\documentclass[prb,epsfig,twocolumn,showpacs,aps,amssymb,superscriptaddress]{revtex4}
\usepackage{bm,graphicx,amsmath}

\begin{document}
\title{Faraday Rotation and Circular Dichroism Spectra of Gold and Silver Nanoparticle Aggregates}

\author{Sahar Pakdel}
\affiliation{Department of Physics, University of Tehran, P. O. Box 14395-547, Tehran, Iran}

\author{MirFaez Miri}
\email{mirfaez_miri@ut.ac.ir}
\affiliation{Department of Physics, University of Tehran, P. O. Box 14395-547, Tehran, Iran}

\begin{abstract}
We study the magneto-optical response of noble metal nanoparticle clusters. We consider the interaction between the light-induced dipoles of particles.
In the presence of a magnetic field, the simplest achiral cluster, a dimer, exhibits circular dichroism (CD).
The CD of a dimer depends on the directions of the magnetic field and the light wave vector. The CD of a populous cluster weakly depends on the magnetic field.
Upon scattering from the cluster, an incident linearly polarized light with polarization azimuth $\varphi$, becomes elliptically polarized.
The polarization azimuth rotation and ellipticity angle variation are sinusoidal functions of $2 \varphi$. The anisotropy and the chirality of the cluster control the
amplitude and offset of these sinusoidal functions. The Faraday rotation and Faraday ellipticity are also sinusoidal functions of $2 \varphi$.
Near the surface plasmon frequency, Faraday rotation and Faraday ellipticity increase.
\end{abstract}




\maketitle

\section{Introduction}\label{intro}


The magneto-optical (MO) response of nanocomposites has attracted a lot of interest.
In view of potential applications in information storage, magnetic field sensors, switches, modulators, optical isolators, circulators, and deflectors\ \cite{book1,book2}, composites containing $\gamma \! \!-\!\!\textrm{Fe}_2\textrm{O}_3$,
$\textrm{Co}\textrm{Fe}_2\textrm{O}_4$, $\textrm{Fe}_3\textrm{O}_4$, Co, and Fe nanoparticles
are studied\ \cite{ziolo,e024441,interface-effect,e174436,Gu,2005,sol-gel}. Recent advances in plasmonics have led to the introduction of {\it magnetoplasmonic} nanostructures composed of ferromagnets and noble metals. The large MO response of magnetoplasmonic composites can be understood heuristically:
The metallic constituent sustains localized or propagating surface plasmons. This gives rise to enhanced electromagnetic field intensity inside the ferromagnetic constituent which exhibits the MO activity.
Magnetoplasmonic crystals\ \cite{nnano201154}, metal films with arrays of holes filled with a magneto-optically active material\ \cite{khanikaev,belo}, Fe/Cu bilayer films\ \cite{ref46supermo.pdf}, Ag/Co/Ag and Au/Fe/Au systems\ \cite{e205120,125132}, Au-coated $\gamma \! \!-\!\!\textrm{Fe}_2\textrm{O}_3$ nanoparticles\ \cite{supermo.pdf,kumar}, and dumbbell-like $\textrm{Ag}\!-\!\!\textrm{Co}\textrm{Fe}_2\textrm{O}_4$ nanoparticle
pairs\ \cite{dumbell,mop2} are thoroughly studied.

Recently, Sep\'{u}lveda {\it et al.}\ \cite{prlasli} have experimentally demonstrated that pure noble metal nanoparticles exhibit measurable
MO response at low magnetic fields (about $ 10^4 \; \mathrm{G}$). This effect was attributed to the Lorentz force induced by
the collective motion of the conduction electrons in the nanoparticle when the localized surface plasmon resonance is excited.
In this experiment, the particles were not in close proximity, thus interparticle coupling effect\ \cite{ghosh} was weak.


In this paper, we study interparticle coupling effect on the MO activity of gold and silver nanoparticle aggregates.
We rely on the celebrated dipole approximation. In this approximation, a particle with diameter much less than the light wavelength is represented as a dipole.
The local electric field acting on any dipole is a superposition of the incident light field and secondary fields produced by other dipoles.
We assume that the aggregate interacts with the external static magnetic field $\bm{B}$ and linearly polarized light specified by the polarization azimuth $\varphi$.
Using the azimuth $\varphi_s$ and ellipticity angle $\mu$ to characterize the vibration ellipse of the field scattered by the aggregate,
we find that the polarization azimuth rotation $\Delta= \varphi_s - \varphi$ and ellipticity angle variation $\mu$ can be expressed as
\begin{eqnarray}
\Delta (\varphi) &=& \Delta_{_\text{str}}(\varphi)+  B \Delta_{_\text{Far}} (\varphi) , \nonumber \\
\mu (\varphi) &=& \mu_{_\text{str}}(\varphi)+  B  \mu_{_\text{Far}} (\varphi). \label{mor}
\end{eqnarray}
The term $\Delta_{_\text{str}}$ ($\mu_{_\text{str}}$) which does not depend on the magnetic field, describes the
{structural}  polarization azimuth rotation (structural ellipticity angle variation), a manifestation of the {\it anisotropy} and {\it chirality} of the aggregate.
The term $B \Delta_{_\text{Far}}$ ($B  \mu_{_\text{Far}}$) which is proportional to the magnetic field, describes the
Faraday rotation (Faraday ellipticity). We find that near the plasmon resonance of the noble metal, both $ \Delta_{_\text{Far}} $ and $\mu_{_\text{Far}}$ increase. We find that
\begin{eqnarray}
\Delta_{_\text{str}}(\varphi)  &\!\! \approx \!\! & \Delta_1 \sin(2 \varphi \!- \!2 \varphi_0) + \Delta_2 , \nonumber \\
\mu_{_\text{str}}(\varphi)  & \! \approx \! & \mu_1 \sin (2 \varphi-2 \varphi'_0)+ \mu_2 , \label{mor2}
\end{eqnarray}
where $\Delta_i $ and $\mu_i $ depend on the light frequency, host medium dielectric constant, nanoparticle radius, etc..
$ \Delta_{_\text{Far}} $ and $\mu_{_\text{Far}}$ are also sinusoidal function of $2 \varphi$.
For the simplest anisotropic and achiral cluster, a {dimer}, we find that $\Delta_2=0 $ and $\mu_2=0$.
For a set of nanoparticles positioned on a helix, $\Delta_2\neq 0 $ and $\mu_2 \neq0$.
For a random gas of particles, which is anisotropic and chiral to some extent, $\Delta_2\neq 0 $ and $\mu_2 \neq0$.
The significance of these results may be seen better by considering optical response of a bulk medium.
For a {\it bulk} medium, the azimuth rotation is also a sinusoidal function of $2 \varphi$.
The amplitude $\Delta_{1}$ and the offset $\Delta_{2}$ are largely controlled by the anisotropy and chirality of
the bulk medium, respectively\ \cite{svirko}.

Optical rotation and circular dichroism (CD) spectra of {bulk} medium are not independent, but are closely connected by Kramers-Kronig relations\ \cite{emeis}.
For metal nanoparticle assemblies, this interrelationship is not yet thoroughly studied.
Drachev {\it et al.}\ \cite{dra} observed large local CD in fractal aggregates of silver nanoparticles
by means of photon scanning tunneling microscopy.
The value of CD averaged over the whole sample was close to zero.
Yannopapas\ \cite{Ya} noted that interacting nanospheres occupying the sites of a rectangular lattice, show a considerable CD around the surface plasmon frequency.
This planar nonchiral structure is indeed extrinsically chiral\ \cite{ex}.
Fan and Govorov\ \cite{govo1,govo2} showed that the helical arrangements of gold nanoparticles
exhibit CD in the visible wavelength range. It is noteworthy that chiral molecules and biomolecules show strong CD in the UV range.
Circular dichroism of metallic nanoparticle assemblies originates in the electromagnetic dipole-dipole interaction between particles.
Recently Kuzyk {\it et al.}\ \cite{govo3} used DNA origami to create left- and right-handed nanohelices
of diameter $34 ~\mathrm{nm}$ and helical pitch $57 ~\mathrm{nm}$, by nine gold nanoparticles of radius $5 ~\mathrm{nm}$.


Inspired by the work of Sep\'{u}lveda {\it et al.} \cite{prlasli}, and Fan and Govorov\ \cite{govo1,govo2},
we study CD of interacting metallic nanoparticles subject to an external static magnetic field.
We find that the simplest {\it achiral} cluster, a {dimer}, exhibits CD in the presence of the magnetic field.
Near the surface plasmon frequency, CD increase considerably.
Quite remarkably, CD depends on the directions of the magnetic field $\bm{B}$ and the light wave vector
$ \bm{k}$ with respect to the dimer axis $\bm{n}$. In particular, CD vanishes if (i) $\bm{k} \parallel \bm{n} $ and $ \bm{B} \perp \bm{n} $,
(ii) $\bm{k} \perp \bm{n} $ and $ \bm{B} \parallel \bm{n} $, and (iii) $\bm{k} \perp \bm{n} $, $ \bm{B} \perp \bm{n} $, and $ \bm{k} \perp \bm{B} $.
This reminds us the intriguing  magnetochiral dichroism\ \cite{md0,md1,md2,md3,barron} of a solution of chiral molecules:
The absorption of a light beam parallel to the magnetic field and a beam antiparallel to the field are slightly different.
Indeed, on the basis of symmetry arguments,
it has been shown\ \cite{portigal} that the dielectric tensor of a chiral medium has a term proportional to $ \bm{k} \cdot \bm{B} $.

Our article is organized as follows. In Sec.\ \ref{model} we introduce the model.
The circular dichroism and Faraday rotation spectra of a dimer, a set of nanoparticles positioned on a helix, and a random gas of nanoparticles
are presented in Secs.\ \ref{Dimer},\ \ref{helix}, and\ \ref{random}, respectively. A summary of our results and discussions are given in
Sec.\ \ref{Discussion}.

\section{Model}\label{model}
\subsection{Polarizability tensor of the nanoparticle }

We adopt the free electron Drude model to describe the magneto-optical response of {\it bulk} metals.
For an electron subject to the static magnetic field $\bm{B}= B (b_x,b_y,b_z)$ and the oscillating electric field
$\bm{E}(t)=\text{Re}[ \bm{E}(\omega) \exp(- \text{i}\omega t) ]$ of the light, the
equation of motion reads
\begin{equation}
m_e \frac{ d^2 \bm{x}}{d t^2} =- m_e \gamma \frac{ d \bm{x}}{d t} -e \Big( \bm{E}(t) +  \frac{1}{c}\frac{ d \bm{x}}{d t} \times \bm{B} \Big). \label{master}
\end{equation}
Here $\bm{x}$, $\gamma$, $m_e$, and $-e <0 $ denote the displacement vector, relaxation constant, effective mass, and charge of the electron, respectively. CGS units are used throughout this paper.

The polarization of the medium is $ \bm{P}(t)=-e n_e  \bm{x}(t) $ where $ n_e $ is the number of conduction electrons per unit volume.
Solving Eq. (\ref{master}) with the ansatz $\bm{x}(t)=\text{Re}[ \bm{x}(\omega) \exp(- \text{i}\omega t) ]$ yields
$\bm{P}(\omega)=-e n_e  \bm{x}(\omega)=\bm{\chi}(\omega) \bm{E}(\omega)$.
For weak magnetic fields (about $ 10^4 \; \mathrm{G}$) the cyclotron frequency
\begin{equation}
\omega_c= \frac{e B}{ m_e  c}
\end{equation}
is much smaller than the light frequency
$\omega$. Thus we calculate the electric susceptibility tensor $\bm{\chi}(\omega)$ and
the permittivity tensor $\bm{\varepsilon}(\omega)=\bm{I}+ 4 \pi \bm{\chi}(\omega)$ to first order in $ \omega_c/\omega$. Indeed
\begin{equation}
\bm{\varepsilon}(\omega)= \begin{pmatrix}  \varepsilon_{ x x}(\omega) &  \text{i} G \omega_c b_z& -\text{i} G \omega_c b_y \\  -\text{i} G \omega_c b_z&  \varepsilon_{ y y}(\omega) &  \text{i} G \omega_c b_x  \\   \text{i} G \omega_c b_y & -\text{i} G \omega_c b_x &  \varepsilon_{ z z}(\omega) \end{pmatrix} ,
\end{equation}
where $\varepsilon_{ x x}=\varepsilon_{ y y}=\varepsilon_{ z z}=1 - { \omega_p^2}/{(\omega^2+ \text{i} \gamma \omega )}$,
$ G={  \omega_p^2  }/{ (\omega^2 + \text{i} \gamma  \omega )^2 }$,  and
$\omega_p= \sqrt{{4 \pi  n_e e^2  }/{m_e}}$ is the plasma frequency.

Now we consider a metal nanoparticle of radius $a$ much smaller than the light wavelength. The polarizability tensor of the spherical nanoparticle is\ \cite{shilova}
\begin{equation}
\bm{ \alpha}(\omega)=a^3 ( \bm{\varepsilon}-\varepsilon_m \bm{I} )  ( \bm{\varepsilon} +2 \varepsilon_m \bm{I} )^{-1},
\end{equation}
where $ \varepsilon_m$ is the dielectric constant of the isotropic host medium and $\bm{I}$ is the identity matrix.
To first order in $ \omega_c/\omega$,
\begin{equation}
\bm{\alpha}(\omega)= \begin{pmatrix}  \alpha_{ x x}(\omega) & - \mathcal{F} \omega_c   b_z& \mathcal{F} \omega_c  b_y \\  \mathcal{F} \omega_c b_z&  \alpha_{ y y}(\omega) & - \mathcal{F} \omega_c b_x  \\  - \mathcal{F} \omega_c b_y & \mathcal{F} \omega_c b_x &  \alpha_{ z z}(\omega) \end{pmatrix} , \label{alphamatrix}
\end{equation}
where $\alpha_{ x x}=\alpha_{ y y}=\alpha_{ z z}= a^3 (\varepsilon_{ x x} - \varepsilon_m) /({\varepsilon}_{ x x} +2 \varepsilon_m ) $ and
\begin{equation}
\mathcal{F} = -3 \text{i} a^3   \frac{  \omega_p^2  }{ \omega (\omega + \text{i} \gamma )^2 }  \frac{\varepsilon_m   }{({\varepsilon}_{ x x} +2 \varepsilon_m )^2}.
\label{factorf}
\end{equation}
The off-diagonal terms of $\bm{\alpha}$ which are proportional to the magnetic field $B$, might seem small at first glance.
However, the factor $\mathcal{F}$ becomes considerably large at frequencies where 
$\text{Re}({\varepsilon}_{ x x} +2 \varepsilon_m )= \pm \text{Im}({\varepsilon}_{ x x}) $. These frequencies are near to the surface plasmon frequency where $\text{Re}({\varepsilon}_{ x x} +2 \varepsilon_m )=0$.

The simple Drude model helps us to understand the origin of plasmon-induced magneto-optical activity of nobel metal clusters\ \cite{prlasli}.
To make a better link to experiments, we use the Drude critical points model\ \cite{Vial and Laroche} for the dielectric function of {\it bulk} Ag and Au.
This model well reproduces the experimental data of Johnson and Christy\ \cite{JC}.
We also modify the dielectric function of {bulk} metal to take into account the finite size effects in
small metallic spheres. We rely on the following radius-dependent dielectric function\ \cite{Hovel}
\begin{eqnarray}
\varepsilon_{ x x}(\omega) &=& \varepsilon^{ \text{bulk}}_{ x x}(\omega)  + \frac{\omega_p^2}{\omega^2+ \text{i} \gamma_{\text{bulk}} \omega}
-\frac{\omega_p^2}{\omega^2+ \text{i} \gamma \omega}, \label{finitesize}
\end{eqnarray}
where $\gamma =\gamma_{\text{bulk}} + {A v_{\text{F}}  }/{ a}$ is the radius-dependent relaxation, $v_{\text{F}}$ is the Fermi velocity, and
the parameter $A$ is of order $1$. To calculate the size corrected $\alpha_{ x x}$ and $\mathcal{F}$, we
assume that $A= 3/4$\ \cite{coronado}, although there is a slight disagreement among authors about $A$\ \cite{Hovel,coronado,Genzel,Vollmer}.
For silver, $ \omega_p=1.386 \times 10^{16} \; \text{s}^{-1}$, $\gamma_{\text{bulk}}=4.584 \times 10^{13} \; \text{s}^{-1}$,
and $v_{\text{F}}=1.39 \times 10^{8}  \; \text{cm}/\text{s}$.
For gold, $ \omega_p=1.32 \times 10^{16} \; \text{s}^{-1}$, $\gamma_{\text{bulk}}= 1.08 \times 10^{14} \; \text{s}^{-1}$,
and $v_{\text{F}}= 1.4 \times 10^{8} \; \text{cm}/\text{s}$.


\subsection{Coupled-dipole equations}

To describe interaction of an incident electromagnetic wave with a set of $N$ identical particles, we use the dipole approximation:
Each metal nanoparticle of radius $a$ behaves as a point dipole with polarizability tensor $\bm{\alpha}$. The local field acting on any dipole is a superposition of the incident field and secondary fields produced by other dipoles.
Thus the coupled-dipole equations (CDE) for the induced dipoles are
\begin{eqnarray}
& &\bm{d}_i(\omega) \! \!=\!\bm{\alpha}(\omega)  \Big(\!\bm{E}(\bm{r}_i,\omega) \!+\!\!\sum_{ j=1}^{N} {
\bm{W}}(\bm{r}_i \!-\!\bm{r}_j) \!\cdot \!\bm{d}_j(\omega) \Big ) . \label{eq2}
\end{eqnarray}
Here $\bm{r}_i$ and $\text{Re}[\bm{d}_i(\omega)  \exp(- \text{i}\omega t) ] $ denote the position and light-induced dipole of the $i$th particle, respectively.
The dyadic Green's function $\bm{W}$ determines the field produced by the oscillating dipole.
\begin{eqnarray}
W_{\alpha,\beta}(\bm{r}_{ij}) \! \!\! & \!=\!& \! \!\! \Big[  (1 \!-\! \text{i} k |\bm{r}_{ij} |)  \Big( \frac{ 3 (\bm{r}_{ij})_{\alpha} (\bm{r}_{ij})_{\beta} - |\bm{r}_{ij} |^{2}   \delta_{\alpha\beta} }{ \varepsilon_m |\bm{r}_{ij} |^{5} } \Big) \nonumber \\
& & \!\! + k^2     \Big( \frac{ |\bm{r}_{ij} |^{2} \delta_{\alpha\beta} - \!  (\bm{r}_{ij})_{\alpha} (\bm{r}_{ij})_{\beta}  }{ \varepsilon_m |\bm{r}_{ij} |^{3} } \Big)  \Big] e^{\text{i} k |\bm{r}_{ij} | },
\end{eqnarray}
where $ \bm{r}_{ij}= \bm{r}_i-\bm{r}_j$, $ k= \omega \sqrt{\varepsilon_m}/c$, and the Greek subscripts denote the Cartesian components.

\section{Dimer}\label{Dimer}
\subsection{Circular dichroism}
It is instructive to consider a dimer, as the simplest {\it achiral} cluster. We assume that the particles
reside at $\bm{r}_1= (0,0,0)$ and $\bm{r}_2=(L,0,0)$. The vector $\bm{n}=(\bm{r}_2-\bm{r}_1)/L$ specifies the axis of the dimer.
The dimer interacts with the electromagnetic field
$ \bm{E}( \bm{r}, \omega) = E_0 \bm{e} \exp( \text{i} \bm{k} \cdot \bm{r} ) $. The polarization vector $ \bm{e}$ and the wave vector
$ \bm{k}=  k ( \sin \theta    \cos \varphi, \sin\theta  \sin\varphi ,\cos\theta) $ are perpendicular.

The light-induced dipole moments obey the equations $\bm{d}_1 \!=\!\bm{\alpha} \Big(\bm{E}(\bm{r}_1,\omega)+{\bm{W}}\cdot \bm{d}_2 \Big ) $ and
$\bm{d}_2 \!=\!\bm{\alpha}  \Big(\bm{E}(\bm{r}_2, \omega)+{\bm{W}}\cdot \bm{d}_1 \Big )$. Here
\begin{eqnarray}
\bm{W} &=& \begin{pmatrix}  \eta + \xi & 0 & 0  \\  0&  \eta &  0  \\   0 & 0& \eta \end{pmatrix} ,\nonumber \\
\eta(L,k) & =&  (k^2- \frac{1}{L^2}+ \frac{\text{i} k}{L})   \frac{e^{\text{i} k L}  }{ \varepsilon_m L},\nonumber \\
\xi(L,k)  & =&(-k^2+\frac{3}{L^2}- \frac{3 \text{i} k}{L}) \frac{e^{\text{i} k L}  }{ \varepsilon_m L}.
\end{eqnarray}
Introducing $  \bm{d}_m=  \bm{\alpha}^{-1} (\bm{d}_1 - \bm{d}_2)$,
$\bm{d}_p= \bm{\alpha}^{-1}  (\bm{d}_1  +\bm{d}_2)$,
$\bm{E}_m = \bm{E}(\bm{r}_1,, \omega)- \bm{E}(\bm{r}_2, \omega) $, and
$\bm{E}_p= \bm{E}(\bm{r}_1, \omega)+ \bm{E}(\bm{r}_2, \omega)$, the coupled-dipole equations can be written as
\begin{eqnarray}
& &(\bm{I} + \bm{W}  \bm{\alpha} ) \bm{d}_m = \bm{E}_m ,\nonumber \\
& & (\bm{I} - \bm{W}  \bm{\alpha} ) \bm{d}_p ~ =\bm{E}_p,
\end{eqnarray}
which are easily solvable. To first order in $ \omega_c/\omega$,
\begin{eqnarray}
\bm{d}_{mx} &\!\!\!\!=\!\!\!\!& \frac{(1 \!\!+\!\! \eta \alpha_{ x x})  E_{m x} \!\!-\!\!(\eta \!\!+\!\! \xi ) \alpha_{ x y} E_{m y}\! \!-\!\!(\eta \!\!+\!\! \xi ) \alpha_{ x z} E_{m z} }{(1+ \eta \alpha_{ x x})(1+ (\eta+ \xi) \alpha_{ x x})} ,\nonumber \\
\bm{d}_{my} &\!\!\!\!=\!\!\!\!& \frac{1}{1 \! \!+\!\! \eta \alpha_{ x x}} \Big(   \frac{\eta \alpha_{ x y}  E_{m x}}{1+ (\eta+ \xi) \alpha_{ x x}} \!+\! E_{m y} \!-\! \frac{ \eta \alpha_{ y z}E_{m z}}{1 \!+\! \eta \alpha_{ x x}}  \Big ) ,\nonumber \\
\bm{d}_{mz} &\!\!\!\!=\!\!\!\!& \frac{1}{1 \! \!+\!\! \eta \alpha_{ x x}} \Big(   \frac{\eta \alpha_{ x z}  E_{m x}}{1+ (\eta+ \xi) \alpha_{ x x}} \!+\!
\frac{ \eta \alpha_{ y z} E_{m y} }{1 \!+\! \eta \alpha_{ x x}} \!+\! E_{m z}  \Big ) ,\nonumber \\
\bm{d}_{px} &\!\!\!\!=\!\!\!\!& \frac{(1 \!\!-\!\! \eta \alpha_{ x x})  E_{p x} \!\!+\!\!(\eta \!\!+\!\! \xi ) \alpha_{ x y} E_{p y}\! \!+\!\!(\eta \!\!+\!\! \xi ) \alpha_{ x z} E_{p z} }{(1- \eta \alpha_{ x x})(1- (\eta+ \xi) \alpha_{ x x})},\nonumber \\
\bm{d}_{py} &\!\!\!\!=\!\!\!\!& \frac{1}{1 \! \!-\!\! \eta \alpha_{ x x}} \Big(   -\frac{\eta \alpha_{ x y}  E_{p x}}{1- (\eta+ \xi) \alpha_{ x x}} \!+\! E_{p y} \!+\! \frac{ \eta \alpha_{ y z}E_{p z}}{1 \!-\! \eta \alpha_{ x x}}  \Big ),\nonumber \\
\bm{d}_{pz} &\!\!\!\!=\!\!\!\!& \frac{1}{1 \! \!-\!\! \eta \alpha_{ x x}} \Big( \! \!-\! \frac{\eta \alpha_{ x z}  E_{p x}}{1- (\eta+ \xi) \alpha_{ x x}} \!-\!
 \frac{ \eta \alpha_{ y z} E_{p y} }{1 \!-\! \eta \alpha_{ x x}} \!+\! E_{p z}  \Big ).  \label{alldmp}
\end{eqnarray}
The absorption cross section of the dimer can be expressed in terms of the auxiliary variables $\bm{d}_m$ and $\bm{d}_p$. Indeed
\begin{eqnarray}
\sigma &=& \frac{4 \pi }{ E_0^2  \sqrt{\varepsilon_m} }   \frac{\omega}{c}
\text{Im}( \bm{\alpha}^{-1 *} \bm{d}_1^* \cdot \bm{d}_1 + \bm{\alpha}^{-1*} \bm{d}_2^* \cdot \bm{d}_2   ),\nonumber \\
&=& \frac{2 \pi }{ E_0^2  \sqrt{\varepsilon_m} }   \frac{\omega}{c}
\text{Im}( \bm{\alpha} \bm{d}_m \cdot \bm{d}_m^*  + \bm{\alpha} \bm{d}_p \cdot \bm{d}_p ^*  ).
\end{eqnarray}


We pay attention to the absorption cross sections $ \sigma_{\pm}$ corresponding to the polarization vectors
\begin{equation}
\bm{e}_{\pm} \!=\! \frac{1}{\sqrt{2}}  ( \cos\theta \cos \varphi \mp \text{i} \sin\varphi,\cos\theta \sin\varphi \pm \text{i} \cos \varphi, -\sin \theta    ).
\label{eplusminus}
\end{equation}
$ \bm{e}_{+}$ and $ \bm{e}_{-}$ describe the right and left circularly polarized light, respectively. A straightforward but lengthy calculation shows that the
dimer is circular dichroic.
\begin{eqnarray}
\sigma_{+}\!\!-\!\!\sigma_{-} \!\!\!&=&\!\!\!   \omega_c
 \Big ( k_x b_x \Lambda_1(\eta,\xi) \!\!+\!\! (k_y b_y \!\!+\!\! k_z b_z) \Lambda_2(\eta,\xi)  \Big) k_x^2 L^2 \nonumber \\
&+ & \!\!\! \omega_c  \Big ( k_x b_x \Lambda_1(-\eta,-\xi) \!\!+\!\! (k_y b_y \!\!+\!\! k_z b_z) \Lambda_2(-\eta,-\xi)  \Big)\nonumber \\
&\times & \!\!\!(4+ k_x^2 L^2),
\end{eqnarray}
where
\begin{eqnarray}
\Lambda_1(\eta,\xi) \!\!  \!\!&=&\!\!\!\! -\frac{4 \pi }{\varepsilon_m \mid \!\! 1 \! \!+\!\! \eta \alpha_{ x x} \!\! \mid^2} \Big[\text{Re}(\mathcal{F}) \!\!+\!\! \text{Im}(  \frac{ 2 \eta\mathcal{F} }{1 \! \!+\!\! \eta \alpha_{ x x}}  )    \text{Im}(\alpha_{ x x} )  \Big ] , \nonumber \\
\Lambda_2(\eta,\!\xi) \!\!  \!\!&=&\!\! \frac{4 \pi }{\varepsilon_m \mid \!\! 1 \! \!+\!\! \eta \alpha_{ x x} \!\! \mid^2}
\Big[\!-\!\text{Re}(  \frac{ 1 \! \!+\!\! \eta \alpha_{ x x}}{  1 \! \!+\!\! (\eta+ \xi) \alpha_{ x x}   }  ) \text{Re}(\mathcal{F})     \nonumber \\
& +&\!\! \!\!  \text{Im}(\!   \frac{  (1 \! \!+\!\! \eta \alpha_{ x x}) ( \eta^* \!\!+\!\!\xi^*)\mathcal{F}^* }{ \mid \!\! 1 \! \!+\!\! (\eta \!\!+\!\! \xi) \alpha_{ x x} \!\! \mid^2  } \!\!-\!\!\frac{ \eta \mathcal{F}}{1 \! \!+\!\! (\eta \!\!+\!\! \xi) \alpha_{ x x} }  ) \text{Im}(\alpha_{ x x} \!) \Big ] .\nonumber \\
\end{eqnarray}

\begin{figure}[t]
\includegraphics[width=0.85\columnwidth]{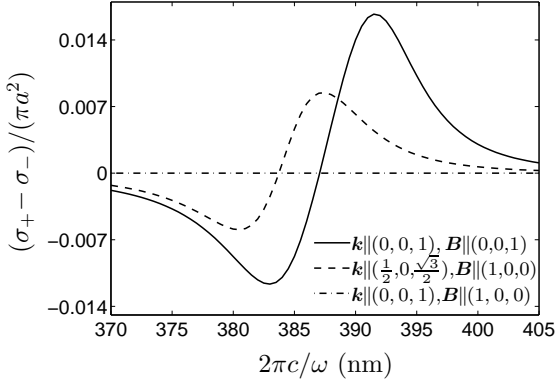}
\caption{$(\sigma_{+} \!-\sigma_{-})/(\pi a^2) $ of an Ag dimer as a function of the wavelength
$2 \pi c/\omega$ for various directions of $\bm{B}$ and $\bm{k}$.
Here $a \!=\! 10~\mathrm{nm}$, $L \!=\! 3a$, $\varepsilon_m \!=\!1.7$ and $ B \!=\! 8000 \; \mathrm{G} $. }
\label{fig1}
\end{figure}

As expected, the achiral dimer exhibits no circular dichroism in the absence of the external magnetic field ($\omega_c =0$).
Quite remarkably, $\sigma_{+}\!-\sigma_{-}$ depends on the directions of the magnetic field $\bm{B} \!=\! B (b_x,b_y,b_z)$ and the wave vector
$ \bm{k} \!=\!(k_x,k_y,k_z)$ with respect to the dimer axis $\bm{n}$, see Fig.\ \ref{fig1}.
The circular dichroism certainly vanishes if (i) $\bm{k} \parallel \bm{n} $ and $ \bm{B} \perp \bm{n} $,
(ii) $\bm{k} \perp \bm{n} $ and $ \bm{B} \parallel \bm{n} $, and (iii) $\bm{k} \perp \bm{n} $, $ \bm{B} \perp \bm{n} $, and $ \bm{k} \perp \bm{B} $.

Note that $ \Lambda_1 \!\propto\! \mathcal{F} $ and $ \Lambda_2 \!\propto\! \mathcal{F} $.
Near the surface plasmon frequency, $\mathcal{F}$ and hence $\sigma_{+}\!-\sigma_{-}$ increase considerably. Moreover, one may expect enhanced circular dichroism where $\text{Re}(1 \pm \eta \alpha_{ x x} ) \!=\!0$ or $\text{Re}(1 \pm (\eta + \xi) \alpha_{ x x} ) \!=\!0$.
We found that these resonances hardly occur for interparticle distance $L> 3 a $. Note that the dipole approximation is not reliable in the region $ L< 3a$.

Figure\ \ref{fig1} exemplifies $\sigma_{+} \!-\sigma_{-}$ of an Ag dimer for various directions of $\bm{B}$ and $\bm{k}$. Here $a \!=\! 10~\mathrm{nm}$, $L=3a$,
and $ B= 8000 \; \mathrm{G}$. For $ \bm{k} \! \! \parallel \! \! (0,0,1) $ and $ \bm{B} \! \! \parallel  \! \! (1,0,0)$ the dimer exhibits no circular dichroism,
but upon light propagation along $\bm{k}  \! \! \parallel  \! \! (0.5,0,0.866) $ the dimer responds quite differently.
Considering the case $ \bm{k} \! \! \parallel \! \! (0,0,1)$ and $ \bm{B} \! \! \parallel \! \! (0,0,1)$, we find that the peaks of
$\sigma_{+} \!-\sigma_{-}$ may increase and shift as the direction of magnetic field varies.

\begin{figure}[t]
\includegraphics[width=0.8\columnwidth]{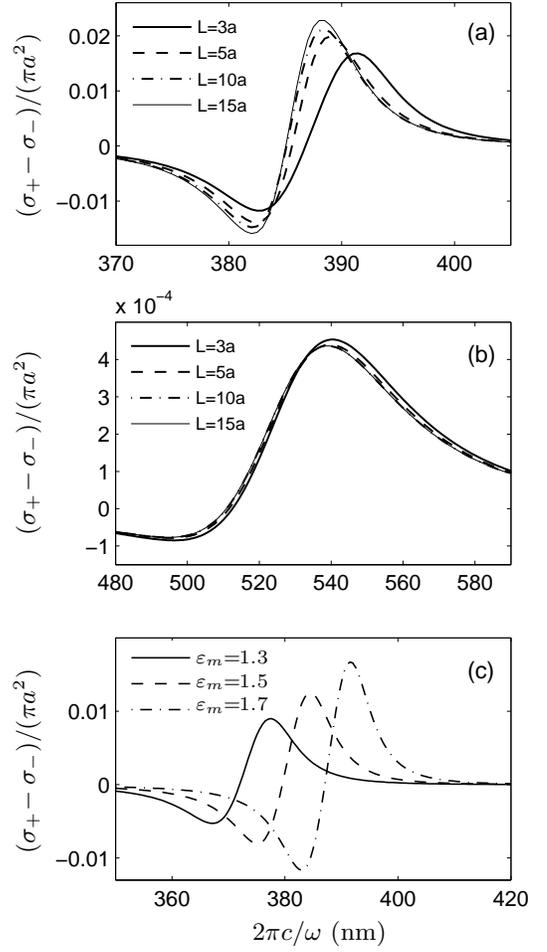}
\caption{$(\sigma_{+} \!-\sigma_{-})/(\pi a^2) $ as a function of the wavelength
$2 \pi c/\omega$ for (a) Ag dimer in a medium with $\varepsilon_m \!=\!1.7$, (b) Au dimer in a medium with $\varepsilon_m \!=\!1.7$, and (c) Ag dimer with interparticle distance  $L \!=\! 3a$.
Here $a \!=\! 10~\mathrm{nm}$, $ \bm{k} \!\parallel \! \hat{\bm{z}} $ and $ \bm{B} \!=\! 8000 \; \mathrm{G} \hat{\bm{z}}$. }
\label{fig2}
\end{figure}

Figure\ \ref{fig2} demonstrates the circular dichroism $\sigma_{+} \!-\sigma_{-}$ in units of the geometric cross section $\pi a^2 $,
as a function of the wavelength $ 2 \pi c/\omega  $. Here $a= 10~\mathrm{nm}$, $ \bm{k} \!\parallel \! \hat{\bm{z}} $ and $ \bm{B}= 8000 \; \mathrm{G} \hat{\bm{z}}$.
An Ag dimer exhibits a considerable negative (positive) dichroism at wavelength $2 \pi c/\omega=383~\mathrm{nm} $ ($391.5~\mathrm{nm}$) when $L=3a$ and $\varepsilon_m \!=\!1.7$, see Fig.\ \ref{fig2}(a). These peaks increase, and slightly shift to lower wavelengths as $L$ increases. The spectrum does not change significantly upon increasing the interparticle distance beyond $L=15a$, since the dipolar interaction becomes quite weak. In the same host medium, the dichroism of an Au dimer reaches its maximum at wavelength $2 \pi c/\omega = 540~\mathrm{nm}$, see Fig.\ \ref{fig2}(b). The dichroism of Ag dimer is about $20$ times stronger than that of Au dimer. This is expected, as $1/\text{Im}({\varepsilon}_{ x x})$ and hence $\mathcal{F}$ of Ag is greater than that of Au. Figures\ \ref{fig2}(c) shows the impact of host medium on the circular dichroism spectrum of Ag dimer. Both peaks of the spectrum grow and shift to higher wavelengths as $\varepsilon_m$ increases.

\subsection{Faraday rotation}

Now we assume that the dimer interacts with a linearly polarized light with the polarization vector
\begin{equation}
\bm{e}=(\cos \varphi ,\sin \varphi, 0) . \label{equ18}
\end{equation}
The wave vector $\bm{k}$ and the
external magnetic field $\bm{B}$ are along the unit vector $\hat{\bm{z}}$.
We measure the electric field $\bm{E}_s$ radiated by the dimer at $ (0,0, h) $, where $ h \gg L>a$.
In the far (radiation) zone the scattered field takes on the form
\begin{eqnarray}
\bm{E}_s &=& (\bm{d}_1  \!- \! \hat{\bm{z}} \hat{\bm{z}} \cdot \bm{d}_1 ) \frac{ k^2 e^{\text{i} k h }  }{ \varepsilon_m   h} \!+\!
(\bm{d}_2 \!-\!  \hat{\bm{z}} \hat{\bm{z}} \cdot \bm{d}_2 ) \frac{ k^2 e^{\text{i} k h }  }{ \varepsilon_m   h} ,\nonumber \\
&=& (  \bm{\alpha} \bm{d}_p -  \hat{\bm{z}} \hat{\bm{z}} \cdot \bm{\alpha} \bm{d}_p  ) \frac{ k^2 e^{\text{i} k h }  }{ \varepsilon_m   h}. \label{takes}
\end{eqnarray}


The electromagnetic wave $\bm{E}_s=  (E_{sx}, E_{sy},0)  $ is elliptically polarized. The vibration ellipse can be characterized by its azimuth $\varphi_s$,
the angle between the semimajor axis and the unit vector $\hat{\bm{x}}$, and its ellipticity $\tan\mu$, the ratio of the length of its semiminor axis to that of its semimajor axis. Operationally defined in terms of measurable intensities,
the Stokes parameters\ \cite{bohren,collet}
\begin{eqnarray}
I &=&  \mid \!  \! E_{sx} \! \! \mid^2 + \mid \! \! E_{sy}  \! \! \mid^2 ,\nonumber \\
Q &=& \mid \!  \! E_{sx} \! \! \mid^2- \mid \! \! E_{sy}  \! \! \mid^2 ,\nonumber \\
U &=&  2  \text{Re}( E_{sx} ^* E_{sy}) ,\nonumber \\
V &=&  2  \text{Im}( E_{sx} ^* E_{sy}),
\end{eqnarray}
determine completely the state of polarization. Indeed
\begin{eqnarray}
& & \varphi_s =\frac{1}{2} \arctan\frac{U}{Q}, \nonumber \\
& &  \mu = \frac{1}{2} \arcsin \frac{V}{I}.
\end{eqnarray}
Note that the azimuth and the ellipticity angle of the linearly polarized incident wave are $\varphi$ and $0$, respectively.
Hereafter we focus on the polarization azimuth rotation $\varphi_s \!-\! \varphi$ and the ellipticity angle variation $\mu$.


\begin{figure}[t]
\includegraphics[width=0.95\columnwidth]{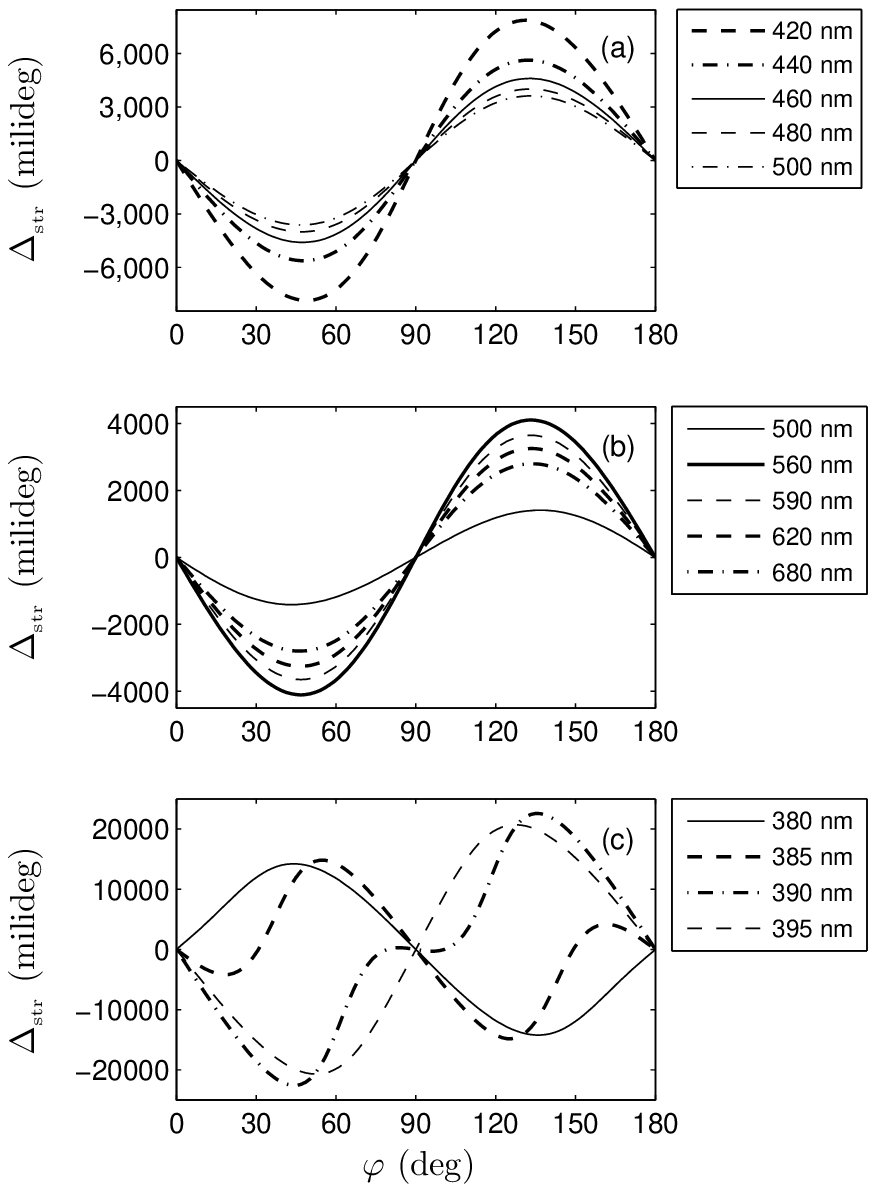}
\caption{ $\Delta_{_\text{str}}(\varphi)$ of (a) Ag dimer, (b) Au dimer, and (c) Ag dimer, as a function of $\varphi$
for various wavelengths $2 \pi c/\omega$. Here $a \!=\! 10~\mathrm{nm}$, $L \!=\! 3a$, $\varepsilon_m \!=\!1.7$, and $B=0$.}
\label{fig3}
\end{figure}

\begin{figure}[t]
\includegraphics[width=0.94\columnwidth]{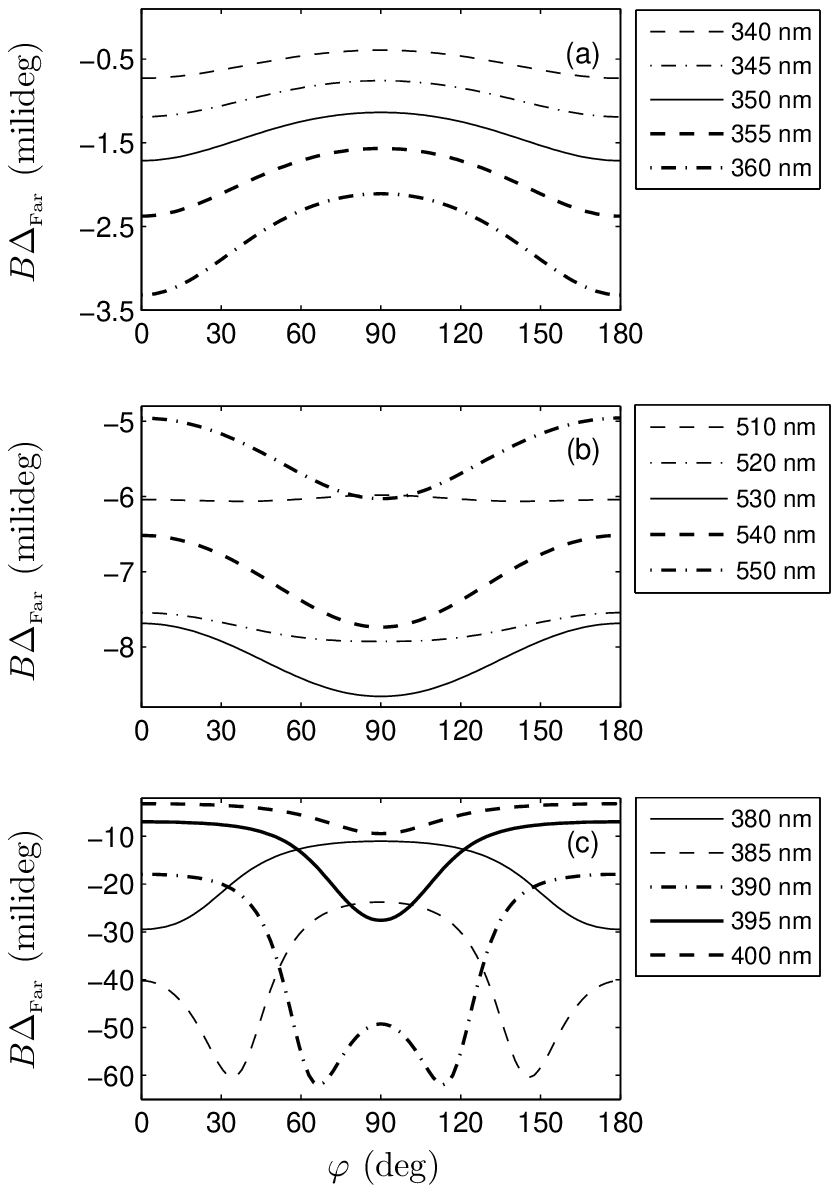}
\caption{ $B \Delta_{_\text{Far}} (\varphi)$ of (a) Ag dimer, (b) Au dimer, and (c) Ag dimer, as a function of $\varphi$
for various wavelengths $2 \pi c/\omega$. Here $a \!=\! 10~\mathrm{nm}$, $L \!=\! 3a$, $\varepsilon_m \!=\!1.7$, and $B= 8000 \; \mathrm{G}$.}
\label{fig4}
\end{figure}

\begin{figure}[t]
\includegraphics[width=0.95\columnwidth]{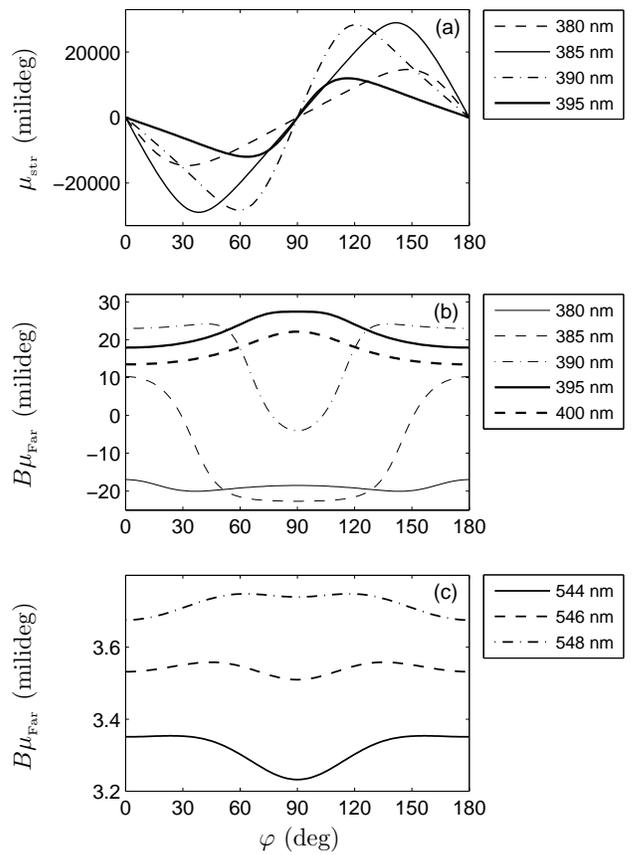}
\caption{ (a)  $\mu_{_\text{str}}(\varphi)$ of Ag dimer, (b) $B  \mu_{_\text{Far}}(\varphi) $ of Ag dimer, and (c) $B  \mu_{_\text{Far}}(\varphi) $ of Au dimer, as a function of $\varphi$ for various wavelengths $2 \pi c/\omega$. Here $a \!=\! 10~\mathrm{nm}$, $L \!=\! 3a$, $\varepsilon_m \!=\!1.7$, and $B= 8000 \; \mathrm{G}$.}
\label{fig5}
\end{figure}

Powered by the analytical expressions (\ref{alldmp}) and (\ref{takes}), we find that to first order in $ \omega_c/\omega$
\begin{eqnarray}
\varphi_s \!-\! \varphi &\!=\!& \frac{1}{2} \arctan(\frac{ J_1 \sin 2 \varphi}{  \cos 2 \varphi \!+\! J_2}) \!-\! \varphi \nonumber \\
&+ & \!\! \!\!\frac{  \omega_c}{2} \frac{ J_1 \! J_3 \sin^2 2 \varphi  \!+\! (J_3 \!\cos 2 \varphi \!+\! \!J_4 )(\cos 2 \varphi \!+\! J_2)}{ (J_1 \sin 2 \varphi)^2+ (\cos 2 \varphi \!+\! J_2)^2 },\nonumber \\
\mu &\!=\!&  \frac{1}{2} \arcsin( \frac{-J_5 \sin 2 \varphi }{ J_2 \cos 2 \varphi \!+\! 1 }) \nonumber \\
&+ & \!\! \!\!\frac{  \omega_c}{2}  \frac{(J_6 \cos 2 \varphi \!+\! \! J_7)( J_2 \cos 2 \varphi \!+ \!\!1 ) \!-\! \! J_4 J_5 \sin^{2} 2 \varphi}
{( J_2 \cos 2 \varphi \!+ \!\!1 )  \sqrt{( J_2 \cos 2 \varphi \!+ \!\!1 )^2 \!-\!(J_5 \sin 2 \varphi)^2}}, \nonumber \\ \label{kolokoli}
\end{eqnarray}
where the parameters $J_1$-$J_7$ are reported in the appendix. As already mentioned in the Introduction,
the polarization azimuth rotation is composed of two terms which are different in nature.
The first term of $\Delta=\varphi_s \!-\! \varphi$ which does not depend on the external magnetic field,
is the {structural polarization azimuth rotation} $\Delta_{_\text{str}}(\varphi)$. Note that a dimer is an anisotropic and achiral object. Switching off the dipolar interaction between the spheres
($\eta, \xi \!\rightarrow \! 0 $), this structural azimuth rotation expectedly vanishes.
The second term of $\varphi_s \!-\! \varphi$ which is proportional to the external magnetic field, is the Faraday rotation $B \Delta_{_\text{Far}} (\varphi)$. Note that $J_3 \propto \mathcal{F}$ and $J_4 \propto \mathcal{F}$, thus the plasmon resonance boosts the Faraday rotation.
Similarly, the first term of $\mu$ describes the structural ellipticity angle variation.
$J_4$, $J_6$ and $J_7$ are proportional to $\mathcal{F}$. The plasmon excitation thus enhances the
second term of $\mu$ which is the Faraday ellipticity.

We have numerically studied $J_1$-$J_7$. We find that in a large domain of parameter space
\begin{equation}
J_1 \! \!\simeq\! \! 1, J_2 \! \!\ll\! \! 1 , J_3 \omega_c  \! \!\ll \! \! 1 , J_4 \omega_c \! \! \ll\! \! 1, J_5 \! \!\ll\! \! 1, J_6 \omega_c \! \! \ll\! \! 1 , J_7 \omega_c \! \! \ll\! \! 1, \label{codition}
\end{equation}
which allows us to rewrite Eq. (\ref{kolokoli}) as
\begin{eqnarray}
\varphi_s \!-\! \varphi & \!\!\simeq\!\! & \!\! -\!\frac{J_2}{2} \sin 2 \varphi
\!+\!   \frac{\omega_c}{2}  \Big( (J_4 \!-\! J_2 J_3) \cos 2 \varphi \!+ \!J_3  \Big) , \nonumber \label{s1}\\
\mu & \!\!\simeq\!\! & -\!\frac{J_5}{2} \sin 2 \varphi + \frac{\omega_c}{2} (-J_7 J_2 \cos 2 \varphi + J_7) .  \label{s3}
\end{eqnarray}
In other words, provided that condition (\ref{codition}) is satisfied, $\Delta_{_\text{str}}$, $B \Delta_{_\text{Far}}$, $\mu_{_\text{str}}$ and $B  \mu_{_\text{Far}}$
are sinusoidal function of $2 \varphi$. These functions can be represented in the form of Eq. ({\ref{mor2}}): For a dimer $\Delta_2=0 $ and $\mu_2=0$.


Now we use Eq. (\ref{kolokoli}) to access the structural polarization azimuth rotation $\Delta_{_\text{str}}$ of a silver dimer as a
function of $\varphi$ and $2 \pi c/\omega$.
We assume that $a \!=\! 10~\mathrm{nm}$, $L \!=\! 3a$, $\varepsilon_m \!=\!1.7$, and $B=0$.
Figure\ \ref{fig3}(a) shows that $\Delta_{_\text{str}}$ is a sinusoidal function of $2 \varphi$.
At wavelength $420 ~\mathrm{nm}$, $\Delta_{_\text{str}}$ has a considerable amplitude $7.84^\circ $.
Upon increasing the wavelength to $500 ~\mathrm{nm}$, the amplitude decreases to $3.62 ^\circ $.
Figure\ \ref{fig3}(b) shows a similar plot for a gold dimer.
Here $\Delta_{_\text{str}}$ is less than $ 4.1 ^\circ $.
Figure\ \ref{fig3}(c) shows that in the narrow window $380 \leqslant 2 \pi c/\omega \leqslant 395 ~\mathrm{nm} $, $\Delta_{_\text{str}}$ of a silver dimer is a nonsinusoidal function. This is because the condition (\ref{codition}) is violated. Quite remarkably,
in this window $\Delta_{_\text{str}}$ can be as large as $ 22.6 ^\circ  $.
Figures\ \ref{fig4}(a) and\ \ref{fig4}(b) show the Faraday rotation of a silver and a gold dimer subject to the magnetic field $B= 8000 \; \mathrm{G}$, respectively.
$B \Delta_{_\text{Far}} (\varphi)$ is a sinusoidal function of $2 \varphi$.
Apparently the maximum of Faraday rotation $< 0.01 ^\circ$ is much less than the maximum of structural azimuth rotation
$ 7.84^\circ $. Figure\ \ref{fig4}(c) shows that in the narrow window $380-400 ~\mathrm{nm} $, $B \Delta_{_\text{Far}} (\varphi)$ of a silver dimer is a nonsinusoidal function.

The structural ellipticity angle variation $\mu_{_\text{str}}$ behaves similar to $\Delta_{_\text{str}}$.
The sinusoidal $\mu_{_\text{str}}$ of a silver (gold) dimer gains its maximum $0.1 ^\circ$ ($3.3 ^\circ$)
at wavelength $480 ~\mathrm{nm}$ ($510 ~\mathrm{nm}$). In the window $380-395 ~\mathrm{nm}$,
$\mu_{_\text{str}}$ of a silver dimer is a nonsinusoidal function, see Fig.\ \ref{fig5}(a).
For the magnetic field $B= 8000 \; \mathrm{G}$ and wavelengths not in the window $330-440 ~\mathrm{nm}$ ($540-556 ~\mathrm{nm}$),
$B  \mu_{_\text{Far}}$ of a silver (gold) dimer is a sinusoidal function of $2 \varphi$.
In this regime, the maximum of Faraday ellipticity angle variation $ 0.004 ^\circ$ ($ 0.0046 ^\circ$) of a
silver (gold) dimer is much less than the maximum of structural ellipticity angle variation
$0.1^\circ$ ($3.3 ^\circ$). The nonsinusoidal behavior of $B  \mu_{_\text{Far}}$ for silver and gold dimer are shown in Figs.\ \ref{fig5}(b) and\ \ref{fig5}(c), respectively.

\section{ Nanoparticles positioned on a helix}\label{helix}

\begin{figure}[t]
\includegraphics[width=0.95\columnwidth]{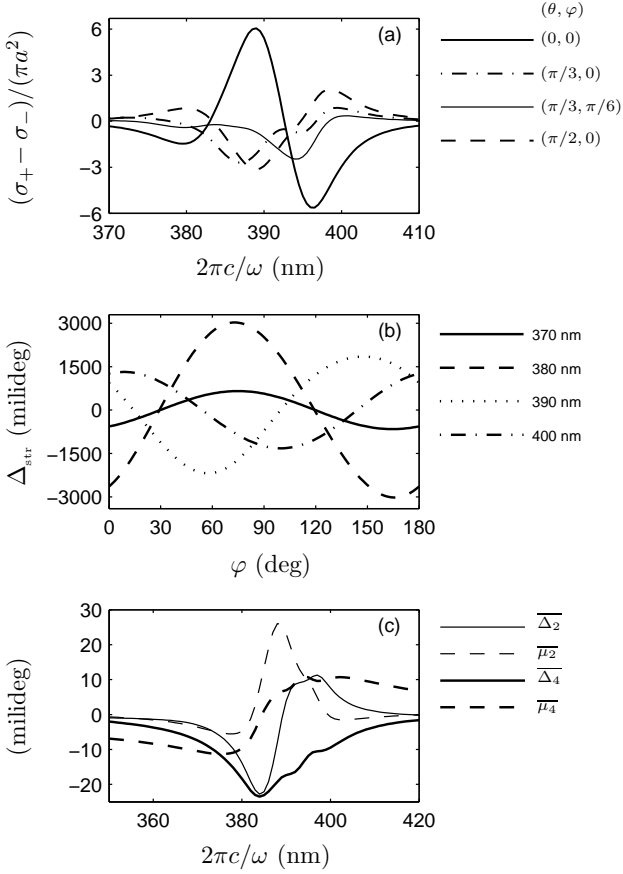}
\caption{ Magneto-optical response of $15$ silver nanoparticles positioned on a helix. $a \!=\! 10~\mathrm{nm}$ and $\varepsilon_m \!=\!1.7$.
(a) $(\sigma_{+} \!-\sigma_{-})/(\pi a^2)$ as a function of $2 \pi c/\omega$, for various directions of $\bm{k}=k (\sin \theta \cos \varphi, \sin\theta \sin\varphi ,\cos\theta)$. Here $B=0$. (b) $\Delta_{_\text{str}}$ as a function of $\varphi$, for various wavelengths $2 \pi c/\omega$.
(c) $\overline{\Delta_2}$, $\overline{\mu_2}$, $\overline{\Delta_4}$, and $\overline{\mu_4}$ as a function of $2 \pi c/\omega$.
Here $B= 8000 \; \mathrm{G}$.
}
\label{fig6}
\end{figure}

\begin{figure}[t]
\includegraphics[width=0.95\columnwidth]{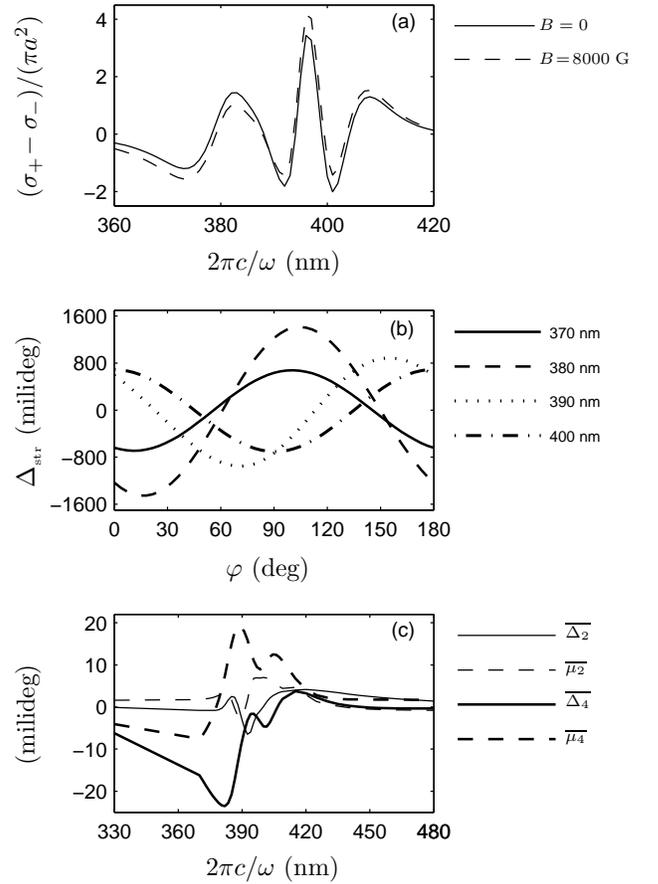}
\caption{ Magneto-optical response of $1000$ silver nanoparticles distributed in an ellipsoid. $a \!=\! 10~\mathrm{nm}$ and $\varepsilon_m \!=\!1.7$.
(a) $(\sigma_{+} \!-\sigma_{-})/(\pi a^2)$ as a function of $2 \pi c/\omega$. (b) $\Delta_{_\text{str}}$ as a function of $\varphi$,
for various wavelengths $2 \pi c/\omega$.
(c) $\overline{\Delta_2}$, $\overline{\mu_2}$, $\overline{\Delta_4}$, and $\overline{\mu_4}$ as a function of $2 \pi c/\omega$.
Here $\bm{k}\! \parallel\! \hat{\bm{z}}$ and $ \bm{B} \!=\! 8000 \; \mathrm{G} \hat{\bm{z}}$.}
\label{fig7}
\end{figure}


We aim to further study correlation between the chirality and the magneto-optical response of a set of nanoparticles. Here
we consider a set of $N$ identical nanoparticles of radius $a$ positioned on a helix. We assume that the $i$th particle resides at
\begin{eqnarray}
x_i &=&R_{\text{helix}} \cos (i \theta_{\text{helix}} \!-\! \theta_{\text{helix}}), \nonumber \\
y_i &=& R_{\text{helix}} \sin (i  \theta_{\text{helix}}  \!- \! \theta_{\text{helix}}) , \nonumber \\
z_i &=&(i-1)  p_{\text{helix}}  \theta_{\text{helix}}    /(2 \pi).
\end{eqnarray}
As an example, we choose $N=15$, $a \!=\! 10~\mathrm{nm}$,
$ R_{\text{helix}}\!=\! 30~\mathrm{nm}$, $ p_{\text{helix}} \!=\! 30~\mathrm{nm}$, and $ \theta_{\text{helix}} =\pi/3$.

We assume that the helical structure interacts with the electromagnetic field $\bm{E}(\bm{r},\omega)=E_0 \bm{e} \exp( \text{i} \bm{k} \cdot \bm{r})$.
We compute the absorption cross sections $\sigma_{\pm}$ corresponding to the polarization vectors $\bm{e}_{\pm}$
introduced in Eq. (\ref{eplusminus}). The dielectric constant of the matrix $\varepsilon_m \!=\!1.7$ and external magnetic field $B=0$.
Figure\ \ref{fig6}(a) delineates $\sigma_{+}\!-\sigma_{-}$ of a helical assembly of silver nanoparticles
as a function of wavelength $2 \pi c/\omega$, for various directions of $\bm{k}=k (\sin \theta \cos \varphi, \sin\theta \sin\varphi ,\cos\theta)$. We can see that the CD spectrum has both positive and negative bands.
The peaks of the CD spectrum change and shift as the direction of light wave vector $\bm{k}$ varies. For light propagation along the helix axis, {\it i.e.}
$ \bm{k} \!\parallel \! \hat{\bm{z}} $, $\sigma_{+}\!-\sigma_{-}$ gains its considerable extrema  $ 6.05\pi a^2$ and $ -5.63\pi a^2$ at wavelengths $389  ~\mathrm{nm}$ and $396  ~\mathrm{nm}$, respectively. We also find that\ \cite{brevity} the CD spectrum does not change significantly as the static
magnetic field increases to $B= 8000 \; \mathrm{G}$.
We observe that the CD spectrum is quite susceptible to neglect of finite size effects in small metallic spheres, see Eq. (\ref{finitesize}).
This point is overlooked in Ref.~\onlinecite{govo1}.

We now assume that $\bm{k} \! \parallel  \! \hat{\bm{z}}$, $\bm{B} \! \parallel \! \hat{\bm{z}} $, and the helical structure
interacts with a linearly polarized light with the polarization vector introduced in Eq. (\ref{equ18}).
Direct numerical solution of the coupled-dipole equations (\ref{eq2}) allows us to obtain the azimuth $\varphi_s$ and ellipticity angle $\mu$ which characterize the vibration ellipse of the scattered field. Figure\ \ref{fig6}(b) shows the structural polarization azimuth rotation
$\Delta_{_\text{str}}(\varphi)$ of a helical structure, as a function of $\varphi$. We observe that $\Delta_{_\text{str}}$ is a sinusoidal function of $2\varphi$. We also find that\ \cite{brevity} $\mu_{_\text{str}}$, $ \Delta_{_\text{Far}}$, and $  \mu_{_\text{Far}}$ are sinusoidal functions of $2\varphi$. Remarkably, for a set of nanoparticles positioned on a helix $\Delta_2\neq 0 $ and $\mu_2 \neq0$, see Eq. ({\ref{mor2}}).

For an arbitrary set of $N$ nanoparticles, we can not provide exact analytical expressions
for $\Delta(\varphi)$ and $\mu(\varphi)$, or numerically explore the whole parameter space.
In other words, the (narrow) windows of parameters where $\Delta(\varphi)$ and $\mu(\varphi)$ are {\it non}sinusoidal function of $2\varphi$, are not known.
However, encouraged by the success of Eq. (\ref{mor2}) in describing the optical behavior of a bulk medium\ \cite{svirko} and a dimer, we propose to measure the well defined quantities
\begin{eqnarray}
\overline{\Delta_2} &=& \int_0^\pi  \Delta_{_\text{str}}(\varphi)   \frac{d \varphi }{\pi} , \nonumber \\
\overline{\mu_2} &=& \int_0^\pi  \mu_{_\text{str}}(\varphi)   \frac{d \varphi }{\pi}, \nonumber \\
\overline{\Delta_4} &=& \int_0^\pi  B \Delta_{_\text{Far}}(\varphi)   \frac{d \varphi }{\pi} , \nonumber \\
\overline{\mu_4} &=& \int_0^\pi   B  \mu_{_\text{Far}}(\varphi)   \frac{d \varphi }{\pi},
\end{eqnarray}
to study correlation between the geometry and the magneto-optical response of a set of nanoparticles.
Figure\ \ref{fig6}(c) shows $\overline{\Delta_2}$, $\overline{\mu_2}$, $\overline{\Delta_4}$ and $\overline{\mu_4}$ of our helical structure as a
function of the wavelength $2 \pi c/\omega$. $\overline{\Delta_2}$ gains its minimum $ -0.023 ^\circ $ and its maximum $ 0.011 ^\circ$ at wavelengths
$384  ~\mathrm{nm}$ and $397  ~\mathrm{nm}$, respectively. $\overline{\mu_2}$ reaches its maximum $ 0.026 ^\circ $ at wavelength $388  ~\mathrm{nm}$.
Note that near the same wavelengths, the circular dichroism spectrum is extremum.
Similar to $\overline{\Delta_2}$, $\overline{\Delta_4}$ has a deep minimum $ -0.023 ^\circ $ at $384  ~\mathrm{nm}$.
However $\overline{\Delta_2}$ and $\overline{\Delta_4}$ are distinct. Notably, $\overline{\Delta_2}$, $\overline{\mu_2}$, $\overline{\Delta_4}$ and $\overline{\mu_4}$
have the same order of magnitude.

\section{Random gas of nanoparticles}\label{random}

We now consider a random arrangement of $N$ identical nanoparticles of radius $a$.
We rely on $N a^3/R_g^3 $ as the volume fraction of the cluster. The radius of gyration $R_g$ which characterizes the size of the system, is defined as
$R_g^2= \frac{1}{N}\sum_{i=1}^{N} \mid \! \bm{r}_i -\bm{r}_{\text{cm}} \!\mid ^2$. Here $\bm{r}_{\text{cm}}$ is
the position vector of the center of mass.

Intuitively, a random gas of nanoparticles is anisotropic and chiral to some extent.
One can quantify chirality of a set of points through the Hausdorff chirality measure\ \cite{buda}.
To characterize the anisotropy of the cluster, one can calculate the moment of inertia tensor
$I_{\alpha,\beta}= \sum_{i=1}^{N} \delta_{\alpha,\beta} \bm{r}_i \cdot \bm{r}_i - (\bm{r}_i)_{\alpha} (\bm{r}_i)_{\beta}$ and its three real eigenvalues
$I_1  \leqslant I_2 \leqslant I_3$. Then an appropriate function of the principal moments of inertia, {\it e.g.}
$I_1/(I_1+I_2+I_3)$ serves as a measure of anisotropy.

We study a set of $1000$ silver nanoparticles of radius $a=10~\mathrm{nm}$, which are distributed in an ellipsoid with semi-axes lengths
$10 \;\mu{\rm m}$, $30 \;\mu{\rm m}$, and $58 \;\mu{\rm m}$.
Deliberately, we did not allowed the particles to be nearer than $3 a$. The volume fraction of the sample is about $0.05$.
We further assume that $\bm{k} \! \parallel  \! \hat{\bm{z}}$, $\bm{B} \! \parallel \! \hat{\bm{z}} $ and $\varepsilon_m \!=\!1.7$.
Our cluster exhibits circular dichroism, see Fig.\ \ref{fig7}(a).
The spectrum has three maxima and three minima, whose position and magnitude are almost independent of $B$.
We observe that\ \cite{brevity} $\Delta_{_\text{str}}$, $\mu_{_\text{str}}$, $ \Delta_{_\text{Far}}$, and $  \mu_{_\text{Far}}$ are
sinusoidal functions of $2\varphi$, see Fig.\ \ref{fig7}(b).
Remarkably, for a random gas of nanoparticles all $\overline{\Delta_2}$, $\overline{\mu_2}$, $\overline{\Delta_4}$ and $\overline{\mu_4}$ are nonzero, and have the same
order of magnitude, see Fig.\ \ref{fig7}(c). As expected, the wavelengths at which $\sigma_{+}\!-\sigma_{-}$, $\overline{\Delta_2}$, $\overline{\mu_2}$, $\overline{\Delta_4}$ and $\overline{\mu_4}$ are extremum, are quite close.

\section{Discussion}\label{Discussion}

On propagation through a {bulk} medium, a linearly polarized electromagnetic wave becomes elliptically polarized.
The polarization azimuth rotation and ellipticity angle variation are due to optical {anisotropy} and optical {activity} of the bulk medium.
The constitutive equation $ 4 \pi P_\alpha= (\varepsilon_{ \alpha  \beta} - \delta_{ \alpha  \beta}) E_\beta + \Gamma_{\alpha \beta \gamma} \nabla_\gamma E_\beta  $
gives the polarization of the bulk medium $\bm{P}$ in terms of the electric field $\bm{E}$. Here
$\varepsilon_{ \alpha  \beta}$ are components of the dielectric tensor describing anisotropy and
$\Gamma_{\alpha \beta \gamma}$ are components of the nonlocality tensor describing optical activity\ \cite{svirko}.
In isotropic optically active media $\varepsilon_{ \alpha  \beta} =n^2 \delta_{ \alpha  \beta}$
and $ \Gamma_{\alpha \beta \gamma}=\Gamma e_{\alpha \beta \gamma}$, where $n$ is the complex refractive index and $e_{\alpha \beta \gamma}$ is the Levi-Civita symbol.
If a left (right) circularly polarized light is incident on the {isotropic} optically active medium, it will propagate as a left (right) circularly polarized light.
The eigenpolarizations of an anisotropic medium with no optical activity ($\Gamma_{\alpha \beta \gamma}=0$) are two linearly polarized waves.
In general, one can measure the polarization azimuth rotation $\Delta$ and ellipticity angle variation $\mu$ as a function of
the azimuth $\varphi$ of an initially polarized wave, to {\it separate} the optical anisotropy and the optical activity of {\it bulk} medium\ \cite{svirko}.

A {\it nanoparticle assembly} is anisotropic and chiral to some extent.
We showed that, to a good approximation, the {structural} polarization azimuth rotation and the structural ellipticity angle variation of a cluster are
sinusoidal functions of $ 2\varphi$, see Eq. ({\ref{mor2}}). Studying the dimer, the helical arrangement and the random gas of nanoparticles, we inferred that
the amplitude $\Delta_{1}$ and the offset $\Delta_{2}$ inform us about the anisotropy and chirality of the cluster, respectively.
Further studies are needed to quantify anisotropy and chirality of the nanoparticle assembly and quantify their link to $\Delta_{1}$ and $\Delta_{2}$.
$\Delta_{1}$ and $\Delta_{2}$ of a random gas of $1000$ silver nanoparticles are of the order of $1 ^\circ $ and $0.01 ^\circ $, respectively,
see Figs.\ \ref{fig7}(b) and\ \ref{fig7}(c). Experimental measurement of these angles is not difficult.

In the presence of an external magnetic field $10000 \; \mathrm{G}$, bulk gold exhibits very weak MO response.
Indeed at wavelength $700~\mathrm{nm}$, $ \varepsilon_{xy}= 10^{-4} + \text{i} 10^{-3} $ for gold, whereas $ \varepsilon_{xy}= 0.8 + \text{i} 0.16 $
for cobalt\ \cite{prlasli}. However, the off-diagonal terms of the polarizability tensor of an isolated metal nanoparticle becomes considerably large
near the surface plasmon frequency\ \cite{prlasli}, see Eqs. (\ref{alphamatrix}) and (\ref{factorf}).
Hui and Stroud studied Faraday rotation of a {\it dilute} suspension of particles in a host.
Relying on the Maxwell-Garnett approximation, they also found that the Faraday effect becomes large near the surface plasmon frequency\ \cite{Hui}.
The effective medium or "mean-field" approximations do not take into account the interaction between particles. Thus we relied on
coupled-dipole equations.
Considering {\it interacting} nanoparticles, we found that Faraday rotation and Faraday ellipticity are also sinusoidal functions of $2\varphi$.
Remarkably, Faraday rotation and Faraday ellipticity of both helical arrangement and random gas of silver nanoparticles are of the order $0.01 ^\circ $, see
Figs.\ \ref{fig6}(c) and\ \ref{fig7}(c).

Recently Kuzyk {\it et al.}\ \cite{govo3} used DNA origami to create left- and right-handed nanohelices
of diameter $34 ~\mathrm{nm}$ and helical pitch $57 ~\mathrm{nm}$, by nine gold nanoparticles of radius $5 ~\mathrm{nm}$.
They even achieved spectral tuning of CD by metal composition: Depositing several nanometers of silver on gold nanoparticles, a blueshift of the CD peak was observed.
In their pioneering work, Fan and Govorov\ \cite{govo1} assumed that helical arrangement of nanoparticles are in a solution and have random orientations.
Thus they performed averaging over the various directions of light wave vector $\bm{k}$ to compute the CD spectrum.
We envisage an ensemble of {\it aligned} helical structures, thus we emphasized the dependence of
the CD spectrum on the direction of $\bm{k}$, see Fig.\ \ref{fig6}(a).
First, we remind that various techniques are developed to align nanorods\ \cite{align1}, {\it e.g.} gold nanorods are incorporated within poly(vinyl alcohol) thin films
and subsequently aligned by heating and stretching the film\ \cite{align2}.
Nanohelices maybe aligned by the same procedures.
Second, we believe that the light pressure can be invoked to generate and align helical arrangement of nanoparticles.
Nahal {\it et al.}\ \cite{nahal} reported generation of spontaneous periodic structures in AgCl-Ag films.
The AgCl layer acts as a waveguide. The incident light scattered by the Ag nanoparticles excites waveguide modes, which interfere with the incident beam.
Silver nanoparticles then immigrate to the minima of the interference pattern.
Electron microphotographs of samples irradiated by a circularly polarized light, suggests the formation of helical arrangement of nanoparticles.
This is further supported by the circular dichroism of the sample, and recent AFM and SEM images\ \cite{nahalrecent}.

We anticipated that the simplest cluster, a dimer, has interesting MO properties. Two dimensional arrangements of dimers can be produced by
electron beam lithography\ \cite{leitner}. Molecular bridges can also be employed to form Au and Ag dimers, trimers, and tetramers\ \cite{feldheim1,feldheim2}.

Our work can be extended in many directions. MO response of trimers, tetramers, two-dimensional arrangements of particles, fractal clusters,
and the influence of particle size dispersion\ \cite{miri1,miri2,miri3,miri4}, are of immediate interest.
We envisage the {\it optical control} of the MO response of a solution
of molecularly bridged dimers. Some molecular bridges, for example azobenzene derivatives, reversibly transform between {\it trans} and {\it cis} conformations upon light radiation. Thus the dimer length and its MO properties can be optically controlled.

\begin{acknowledgments}
We would like to thank A. Nahal from University of Tehran, for fruitful discussions.
\end{acknowledgments}

\appendix
\section{The parameters $J_1$-$J_7$}\label{app1}

The parameters $J_1$-$J_7$ introduced in Eq. (\ref{kolokoli}) can be written as

\begin{eqnarray}
J_0 & \!\!=\!\!&  \mid \!\!  \alpha_{ x x} \!-\! \eta \alpha_{ x x} ^2
\!\! \mid^2 + \mid \!\!  \alpha_{ x x}- (\eta+ \xi) \alpha_{ x x} ^2 \!\! \mid^2 ,\nonumber \\
J_1 & \!\!=\!\!&    \frac{2}{J_0} {  \text{Re} [(\alpha_{ x x} \!-\! \eta \alpha_{ x x}^2 )  (\alpha_{ x x}^*- (\eta^*+ \xi^{*}) \alpha_{ x x}^{*2})] } , \nonumber \\
J_2 &\!\!=\!\!&  \frac{1}{J_0} ({\mid \!\!  \alpha_{ x x} \!-\! \eta \alpha_{ x x} ^2   \!\! \mid^2 - \mid \!\!  \alpha_{ x x}- (\eta+ \xi) \alpha_{ x x} ^2 \!\! \mid^2} )  , \nonumber \\
J_3 &\!\!=\!\!&  \frac{2}{J_0}  { \text{Re}[(\alpha_{ x x} \!-\! \eta \alpha_{ x x} ^2 ) \mathcal{F}^*  \!+\!(  \alpha_{ x x}^*- (\eta^*+ \xi^*) \alpha_{ x x} ^{*2}) \mathcal{F} ]}     , \nonumber \\
J_4 &\!\!=\!\!&    \frac{2}{J_0}  {\text{Re}[ (\alpha_{ x x} \!-\! \eta \alpha_{ x x} ^2)\mathcal{F}^*  \!-\! (  \alpha_{ x x}^* \!-\! (\eta^* \!+\! \xi^*) \alpha_{ x x} ^{*2}) \mathcal{F} ]}  ,\nonumber \\
J_5 & \!\!=\!\!&    \frac{2}{J_0} {  \text{Im} [(\alpha_{ x x} \!-\! \eta \alpha_{ x x}^2 )  (\alpha_{ x x}^*- (\eta^*+ \xi^{*}) \alpha_{ x x}^{*2})] } , \nonumber \\
J_6 &\!\!=\!\!&  \frac{-2}{J_0}  { \text{Im}[(\alpha_{ x x} \!-\! \eta \alpha_{ x x} ^2 ) \mathcal{F}^*  \!+\!(  \alpha_{ x x}^*- (\eta^*+ \xi^*) \alpha_{ x x} ^{*2}) \mathcal{F} ]}     , \nonumber \\
J_7 &\!\!=\!\!& \!\!   \frac{-2}{J_0}  {\text{Im}[ (\alpha_{ x x} \!-\! \eta \alpha_{ x x} ^2)\mathcal{F}^*  \!\!-\!\! (  \alpha_{ x x}^* \!-\! (\eta^* \!\!+\! \xi^*) \alpha_{ x x} ^{*2}) \mathcal{F} ]}   .
\end{eqnarray}


\begin{thebibliography}{99}
\bibitem{brevity}
For the sake of brevity, data are not shown here.
\end{thebibliography}
\end{document}